# Experimental realization of Energy modulation of high-order R-TEM laser modes in Radially polarized cylindrical vector beam

Brijesh Kumar Mishra, and Brijesh Kumar Singh

*Department of Physics, School of Physical Sciences, Central University of Rajasthan, Ajmer 305817, Rajasthan, India*
*brijeshsingh@curaj.ac.in*

A *In this work, an experimental approach is introduced to redistribute optical energy among the multiple concentric core rings of high-order R-TEM laser modes, differing from conventional high-order R-TEM modes that inherently exhibit non-uniform energy across their rings. By employing a diffractive optical element formed from a binary phase mask with two oppositely phased regions, the energy sharing between the rings can be tuned to achieve a variable intensity ratio in the ring pattern. The resulting modulated high-order R-TEM modes are expected to surpass standard R-TEM modes for applications requiring ring structures with nearly equal intensity, such as micro- and nanoparticle manipulation, optical lithography, and near-field optical data storage.*

**Introduction**

Most laser systems exhibit a uniform polarization across the beam aperture and are typically linearly polarized. Any deviation from this uniformity such as the introduction of ellipticity, rotation, or other instrumental effects can significantly influence system performance [1]. Beams with a non-uniform polarization distribution across the beam cross-section are known as vector beams. A notable class of vector beams is the cylindrical vector (CV) beam, in which the state of polarization (SOP) varies spatially while maintaining cylindrical symmetry. Radial and azimuthal polarizations represent two orthogonal and fundamental forms of CV beams, and other CV polarization states can be generated through suitable superpositions of these two basic modes [2]. Radially polarized beams under a low numerical aperture focusing system show a cylindrically symmetrical profile in terms of field intensity and polarization distribution [3–5]. Radially polarized beams were first generated using a conical optical element inside a laser oscillator [6]. Analytical studies show that the fundamental R-TEM$_{01}$ mode of a radially polarized beam shows axial optics forces better than a linearly polarized beam on the Rayleigh particle [7]. In addition, higher-order radially polarized Laguerre-Gaussian beams provide better performance than lower-order mode and can produce focal spots smaller than the Abbe limit of $0.5\lambda$ [8–10].

Optical trapping and manipulation of micro- and nanoparticles [11], acceleration of charged particles [12], nonlinear optics [13], microscopy [14,15] near-field optics [16], quantum information [17], and the improvement of signal quality in free-space optical communications [18] are of the numerous applications in which CV beams of different orders have shown notable influence. Most of the research on CV beams, whether theoretical or experimental, has focused on fundamental mode, which is a doughnut-shaped beam with a uniform intensity profile [19]. A high-order CV beam, having two or more rings expected to have exceptional focusing qualities to generate a smaller spot size and long depth of focus as well as dual focus [20] and a dark focal spot [21] at the focus under tight focusing conditions, according to theoretical research and numerical simulations [8,22].

There are numerous approaches to producing CVBs, including interferometric methods [23], spatial light modulators (SLMs) [24,25], as well as other strategies that typically require intricate experimental arrangements. The Nd:YAG laser cavity was equipped with a specifically made photonic crystal mirror for the stable creation of high-order modes [26,27]. The primary methods include transforming the polarization of the initial beam using subwavelength gratings [28,29], as well as employing crystalline and thin-film [19]. In addition to this intra-cavity approach, several extra-cavity techniques have been introduced for generating double-ring-shaped beams. In one such method, a uniaxial Iceland spar crystal is employed to produce higher-order CV beams [30–32]. Nevertheless, this has no influence over the order of the generated beam, which is not defined. Another method creates a double-ring-shaped radially polarized spiral beam by combining a binary axicon with an interference polarizer [33,34]. This technique requires a complex experimental arrangement, and the resulting double rings are not clearly separated. Moreover, it exhibits deviations from axial symmetry and lacks azimuthal uniformity in the ring structure. Additionally, by modifying the Pancharatnam-Berry phase [29,35], which is connected to the local polarization changes of light, a metasurface was employed to transform a linearly polarized Gaussian beam into a double-ring-shaped CV beam.

Still, these studies do not demonstrate how the optical energy shifts among the produced rings, even though the focus should be on generating high-intensity rings that remain stable compared to the low-intensity rings of high-order R-TEM modes of RPCV beam, which will be crucial for tasks such as optical trapping and lithography [36,37]. However, as the mode order increases, the number of rings also increases, but these outer rings have lower intensity compared to the central ring, which has high intensity. Because of this uneven intensity distribution, it is not possible to use all the concentric rings and lobes with the same level of effectiveness. Therefore, having continuous control over how the energy is shared between the rings of a high-order R-TEM modes of RPCV beam and, for this, energy modulation is a technique for boosting the capabilities of R-TEM modes [38] will be a useful tool for multiple micro- and nanoparticle trapping in optical tweezers, as well as for increasing resolution, improving pattern fidelity, maximizing exposure, and enabling accurate alignment in lithographic procedures, and is also important for achieving the best results in many practical applications. For this reason, the ability to generate a multiple-ring CV beams with adjustable energy balance among the rings is important for achieving better performance across various applications.

In this work, we experimentally demonstrate controllable energy modulation among the multiple concentric rings of high-order R-TEM modes. This allows us to create evenly high-intensity and opposite intensity distributions, with the ability to stably adjust the intensity between the rings in a controlled manner. Thus, the modulated R-TEM modes of the CV beam gain the capability to use several concentric rings simultaneously with similar effectiveness in various applications. This approach can also be applied to controlling structured scalar light fields [39,40]. For our purpose, we use a binary phase mask consisting of two regions with opposite phases. This mask was numerically simulated, and for the experimental part we use a computer-generated hologram, displayed on the SLM and modulates laser beam is further converted into R-TEM modes using S-waveplate. The advantage of this technique is that both the amplitude and phase of the wave functions can be confined to the first diffraction order by modulating only the phase of the beam, making it easy to implement using a spatial light modulator (SLM). Further, we focus only on the transformation of high-order R-TEM modes of CV beams in this work.

### Theory

To study the effect of radial and longitudinal components on the modulation of optical energy, we used vector diffraction theory. On the basis of this theory Youngworth and Brown calculated the intensity distributions of cylindrical vector beams with azimuthal or radial polarization in the focal region. The electric field in the focal region of a radially polarized beam focused by an aplanatic lens is separated into two components: the axial (longitudinal) field and the radial (transverse) field. [41] as

$$\mathbf{e} = e_\rho \hat{e}_\rho + e_z \hat{e}_z \quad (1)$$

Where, $\hat{e}_\rho$ and $\hat{e}_z$ are the unit vectors in the radial and beam propagation directions, respectively. For a purely radially polarized beam, there is no azimuthal component through focusing on image space. As described by Youngworth and Brown, the electric field vector (radial component $e_\rho$ and longitudinal component $e_z$) near focus can be expressed in cylindrical dimensions as: [5]

$$e_\rho = -\frac{iA}{\pi} \int_0^\alpha \int_0^{2\pi} \cos^{1/2}\theta \sin\theta \cos\theta \, \cos(\phi - \phi_s)$$
$$l_0(\theta) e^{ik(z\cos\theta + \rho\sin\theta\cos(\phi - \phi_s))} \, d\phi \, d\theta \quad (2)$$

$$e_z = -\frac{iA}{\pi} \int_0^\alpha \int_0^{2\pi} \cos^{1/2}\sin^2\theta \, l_0(\theta)$$
$$e^{ik(z\cos\theta + \rho\sin\theta\cos(\phi - \phi_s))} \, d\phi \, d\theta \quad (3)$$

The identity can be used to perform the integrations over $\varphi$:

$$\int_0^{2\pi} \cos(n\varphi) e^{(ik\rho\sin(\theta)\cos\varphi)} \, d\varphi = 2\pi i^n J_n(k\rho\sin(\theta)) \quad (4)$$

where $J_n(k\rho \sin(\theta))$ denotes a Bessel function of the first kind, of order $n$. The electric fields near focus take on the following form when a pupil is illuminated by radial polarization. [8]

$$e_\rho(\rho,z) = A \int_0^\alpha \cos^{1/2}\sin(2\theta) l_0(\theta) J_1(k\rho \sin\theta) e^{ikz \cos\theta} d\theta \quad (5)$$

$$e_z(\rho,z) = 2iA \int_0^\alpha \cos^{1/2}\sin^2\theta \, l_0(\theta) J_0(k\rho \sin\theta) e^{ikz \cos\theta} dZ \quad (6)$$

Like paraxial solutions, it is easily confirmed that a radially illuminated objective generates a focal region with an on-axis null at all axial distances z from the paraxial focus. We express the field components using cylindrical coordinates $(\rho, \varphi, z)$, where the beam propagates in the positive $z$ direction and focused on the focal plane ($z = 0$) and origin is taken as the geometric focal point. To ensure that the truncated focusing beam retains the same optical power in the image space, the power coefficient (A) in Eqs. (5) and (6) are set to 1. The relative amplitude of the electric field at the pupil plane is defined by the apodization function $l_0(\theta)$. The maximum focusing angle $\alpha$ is given by $\sin^{-1}(NA/n)$, where $NA$ is the numerical aperture, and $k$ is the wavenumber in the image space, given by $k = 2\pi/\lambda$.

To calculate Eqs. (5) and (6), we must use the relative amplitude $l_0(\theta)$ corresponding to R-TEM$_{pl}$ mode. A Bessel-Gaussian beam, derived as a paraxial solution of Maxwell's vector wave equation, was employed to generate a single doughnut-shaped radially polarized beam, taking the form $l_0(\theta)$ for the R-TEM$_{pl}$ modes,

$$l_0(\theta) = \left(\frac{\beta_0^2 \sin(\theta)}{\sin^2(\alpha)}\right) exp\left(-\frac{\beta_0^2 \sin^2(\theta)}{\sin^2(\alpha)}\right) L_p^l\left(\frac{2\beta_0^2 \sin^2(\theta)}{\sin^2(\alpha)}\right) \quad (7)$$

Where, the generalized Laguerre polynomial is represented in this context by $L_p^l$ and $\beta_0$ is the ratio of the aperture radius to the beam waist and is known as the truncation parameter. $\beta_0$ should be larger than 1 because the outer ring of the R-TEM$_{11}$ beam will be completely blocked by the pupil if $\beta_0 < 1$.

To achieve the modulation of optical energy among the different concentric rings with a variable intensity ratio among the rings of the high order R-TEM$_{pl}$ modes of a CV beam we modulate its phase distribution by employing a binary phase mask, of two regions of opposite phases (0 and π) in the cross-section of the propagating beam. This is accomplished, both experimentally and numerically, by multiplying Eq. (1) with the phase mask function $\phi_{mask}(\theta)$, as defined in Eq. (8). This operation introduces a wavefront modulation that consequently affects the focal intensity distribution.

$$\varphi_{mask}(\theta) \propto \begin{cases} -1 & \theta \leq \eta \cdot \alpha \\ 1 & \eta \cdot \alpha < \theta \leq \alpha \end{cases} \quad (8)$$

where, $0 \leq \eta \cdot \alpha \leq \alpha$, $\alpha$ denotes the maximum focusing angle determined by the numerical aperture (NA) of the lens, $\theta$ represents the angular coordinate, and $\eta$ is a dimensionless parameter. The modified intensity distribution in the focal plane is numerically computed by tuning $\eta$, enabling the characterization of energy modulation. In this study, we set the numerical aperture = 0.5, the refractive index $n = 1$, and the wavelength to $\lambda = 532$ nm. The length is normalized in units of wavelength.

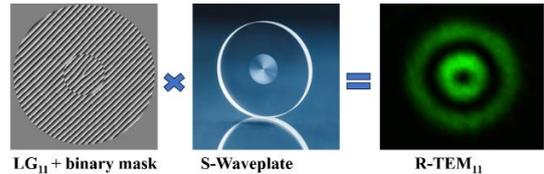

*Figure 1* Conventional setup for realizing modulated R-TEM laser modes at the Fourier plane.

The experimental setup was used to achieve the modulation of optical energy of the high order R-TEM$_{pl}$ modes of CV beam experimentally shown in Figure 1. Nd:YAG diode laser operating at 532 nm works as a source. A polarizer (P), a biconvex lens $L$(f = 20 cm) and a circular aperture (CA), respectively, are used to collimate the beam and produce a linearly polarized Gaussian beam of fixed diameter. A computer-generated hologram (CGH) is employed to transform this linearly polarized Gaussian beam into a high-order Laguerre-Gaussian beam (LGB). The design of the CGH depends on the desired mode order, with different hologram patterns used to generate different LG modes. In a cylindrical coordinate framework, the paraxial solutions of the vector

Helmholtz equation are described by Laguerre-Gaussian modes. By considering the standard beam parameters and applying the paraxial approximation, the CV beam corresponding to the $TM_{0n}$ mode can be written as: [42]

$$CV(\rho, z) = C_0 \frac{\omega_0}{\omega} \left(\frac{\rho\sqrt{2}}{\omega}\right) L^1_{n-1}! \left(\frac{2\rho^2}{\omega^2}\right) \exp\left(-\frac{\rho^2}{\omega^2}\right)$$
$$\exp\left(-\frac{ik\rho^2}{2R} - ikz + i\,2n\,\varphi(z)\right) \quad (9)$$

where $\rho$ is the radial coordinate, $C_0$ is a constant, $\omega_0$ is the beam waist at $z = 0$, and $z$ is the propagation distance from the waist. The beam radius is given by $\omega(z) = \omega_0\sqrt{1 + (z/z_0)^2}$, where $z_0 = k\omega_0^2/2$ is the Rayleigh range. The radius of curvature is $R(z) = (z_0^2 + z^2)/z$, and the Gouy phase is $\phi(z) = \arctan(z/z_0)$, such that $2n(z)$ represents the Gouy phase shift. Here, $k$ denotes the wave number, and $L_{n-1}$ is the generalized Laguerre polynomial of first order and degree $n - 1$. The complex amplitude of the fundamental CV $TM_{01}$ mode in Eq. (10) is equivalent to that of a Laguerre-Gaussian $LG_{01}$ mode [$n = 1$ in Eq. (9)]. LG modes are complete and orthogonal set of solutions for the paraxial wave equation [43]. The complex amplitude of a helical-type $LG_p^l$ mode, with radial and azimuthal indices $p$ and $l$, are usually described as [9]

$$LG_{pl}(\rho, \varphi, z)$$
$$= \frac{\omega_0}{\omega}\left(\frac{\rho\sqrt{2}}{\omega}\right)^{|l|} L_p^{|l|}\left(\frac{2\rho^2}{\omega^2}\right) \exp\left(-\frac{\rho^2}{\omega^2}\right)$$
$$\exp\left(-\frac{ik\rho^2}{2R} - ikz + i(2p + |l| + 1)\varphi(z)\right)$$
$$\exp(il\varphi) \quad (10)$$

Where, $L_p^l$ is generalized Laguerre polynomial of degree $p$ and order $l$, $(2p + |l| + 1)$. $\varphi(z)$ represents the Gouy phase and $\varphi$ denotes the azimuthal angle. All other terms remain the same as described for Eq. (4.9). The Gouy phase and the vortex phase $\exp(il\varphi)$ both differ in Eq. (9) and Eq. (10). After generating the high-order LG modes, we applied the binary phase mask function $\phi_{mask}(\theta)$ to the CGH used for LG mode generation and displayed the updated CGH on the SLM (Holoeye LC-R 720), which has an active area of $1280 \times 768$ pixels (pixel size: 19 μm). After modulating the phase of the high-order LG beam with the updated CGH, an S-waveplate (SWP) is used to convert the LG beam into a high-order R-$TEM_{pl}$ modes of RPCV beam. It is important to note that any polarization state between radial and azimuthal (including hybrid states) can be generated using the SWP by rotating its optical axis with respect to the polarizer. Finally, the generated high-order R-$TEM_{pl}$ mode is subsequently focused by a biconvex lens $L$(f = 30 cm) and captured using an optical camera (Thorlabs DCC1645C; pixel size: 3.6 μm, resolution: $1280 \times 1024$). A neutral density filter (NDF) is inserted in the detection path to prevent camera saturation (shown in the Fig 4.2).

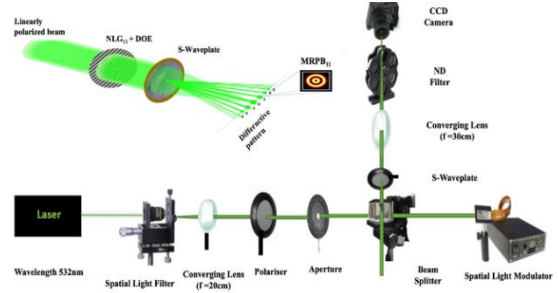

*Figure 2 Experimental setup for generating radially polarized beam and energy modulation distribution*

**Result**

First, we discuss the energy modulation among the rings of the high-order R-$TEM_{11}$, R-$TEM_{21}$, R-$TEM_{31}$ modes of the radially polarized CV beam having non-uniform peak intensity distributions between the central lobe and outer rings, as shown in the Figure 3 (a, d), Figure 4 (a, d), and Figure 5 (a, d). After modulation of the phase of the input RPCV beam the resultant two-dimensional (2D) modulated intensity profile of standard laser modes generated at the focal plane is shown in Figure 3 (b, c, e, f), Figure 4 (b, c, e, f), and Figure 5 (b, c, e, f), respectively. In all Figure the first row (a, b, c) shows simulated two- dimensional (2D) intensity profile, the second row (d, e, f) shows the experimental 2D profiles. The 1D profile of total field components is plotted in white under the 2D profile respectively. All the results shown in a particular row correspond to a particular value of η mentioned in that row. The scale bar shown in the Figure (a) of each Figure 3-5 is also valid for all corresponding Figs. of Figure 3, Figure 4 Figure 5.

To modulate the intensity distribution in all three high-order R-$TEM_{11}$, R-$TEM_{21}$, and R-$TEM_{31}$ modes of the radially polarized CV, we tune the tunning parameter η from 0 to 1 for getting the desired result. We observe that when we continuously increase the value of η from 0 to 1, we are getting the shuffling of optical energy between the central lobe and the outer ring. Further, in continuation of energy modulation, first we tested R-TEM$_{11}$ mode. For η = 0.45, the outer low- intensity ring becomes highly intense by enhancing its peak intensity by about 5.71 times. The central highly intense ring becomes a low-intensity ring after reducing its peak intensity by about 3.41 times with respect to the normal mode peak intensity [Figure 3(b)]. Thus, the 2D transverse intensity profile of the modulated R-TEM$_{11}$ mode in Figure 3(b) is just the reverse of the standard intensity profile of the R-TEM$_{11}$ mode in Figure 3(a). Further, for η = 0.54, the peak intensity of the central low-intensity ring is now enhanced by 3.25 times. The peak intensity of the outer ring remains unchanged with respect to Figure 3(b). The central ring and outer ring both have an equal peak intensity distribution, as shown in the first row of Figure 3(c).

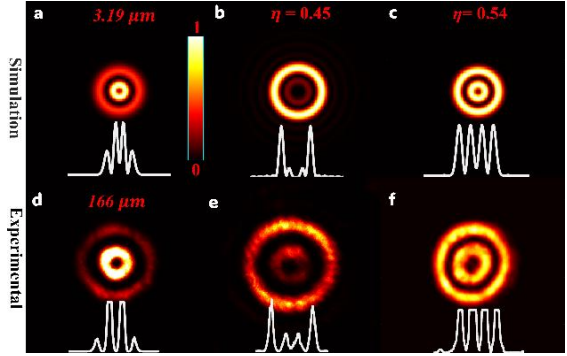
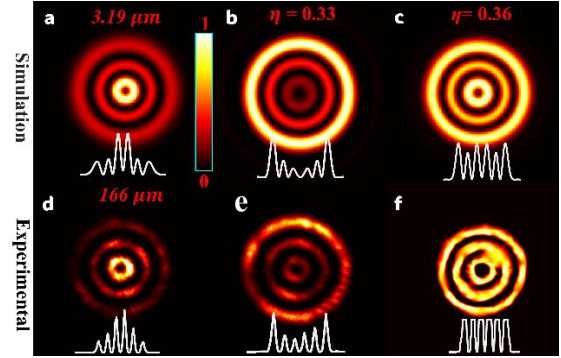

*Figure 3* The first row presents the simulated high-order R-TEM$_{11}$ mode profiles, while the second row shows the corresponding experimental results. The first column displays the unmodulated beam profile, whereas the second and third columns show the modulated profiles obtained at η = 0.45 and η = 0.55, respectively, demonstrating opposite and equal intensity redistribution among the rings. The scale and colour bar provided in Figure 3 (a, d) apply to the corresponding panels.

*Figure 4* The first row presents the simulated high-order R-TEM$_{31}$ mode profiles, while the second row shows the corresponding experimental results. The first column displays the unmodulated beam profile, whereas the second and third columns show the modulated profiles obtained at η = 0.45 and η = 0.55, respectively, demonstrating opposite and equal intensity redistribution among the rings. The scale and color bar provided in Figure 4 (a, d) apply to the corresponding panels.

Similarly, after testing the energy modulation for R-TEM$_{11}$ mode, we obtained similar results for R-TEM$_{21}$ mode and R-TEM$_{31}$ mode following the same trend as in R-TEM$_{11}$ mode. As in the case of R-TEM$_{21}$ mode at η = 0.33, we obtained an opposite intensity profile compared to the standard intensity profile shown in Figure 4(a), where the outer low-intensity ring becomes highly intense after enhancing its peak intensity by about 4.52 times. The central highly intense ring becomes a low-intensity ring by reducing its peak intensity by about 3.21 times with respect to the normal mode peak intensity [Figure 4(b)]. Thus, the 2D transverse intensity profile of the modulated R-TEM$_{21}$ mode in Figure 4(b) is just the reverse of the standard intensity profile of the R-TEM$_{21}$ mode in Figure 4(a). Further, for η = 0.36, the peak intensity of the central low-intensity lobe is now enhanced by 3.25 times. The peak intensity of the outer ring remains unchanged with respect to Figure 4(b). The central lobe and outer ring both have an equal peak intensity distribution, as shown in the first row of Figure 4(c).

Similarly, in case of R-TEM$_{31}$ mode, we obtained an opposite intensity profile Figure 5(b) at η = 0.56, where the outer low-intensity ring becomes highly intense after enhancing its peak intensity by about 5.12 times. The central highly intense ring becomes a low-intensity ring after reducing its peak intensity by about 4.01 times with respect to the normal mode peak intensity Figure 5(a). Thus, the 2D transverse intensity profile of the modulated R-TEM$_{31}$ mode in Figure 5(b) is just the reverse of the standard intensity profile of the R-TEM$_{31}$ mode in Figure 5(a). Also, for η = 0.62, the peak intensity of the central low-intensity ring is now enhanced by 3.25 times. The peak intensity of the outer ring remains unchanged with respect to Figure 5(b). The central lobe and outer ring both have an equal peak intensity distribution, as shown in the first row of Figure 5(c).

**Discussion**

To achieve the desired beam modulation, a linearly polarized LG beam was first generated and then modulated with a grating function and a binary phase mask. This modified LG beam was subsequently passed through an S-waveplate, which converted the modulated input beam into a cylindrical vector (CV) beam.

The simulation and experimental results discussed that the energy modulation of the R-TEM

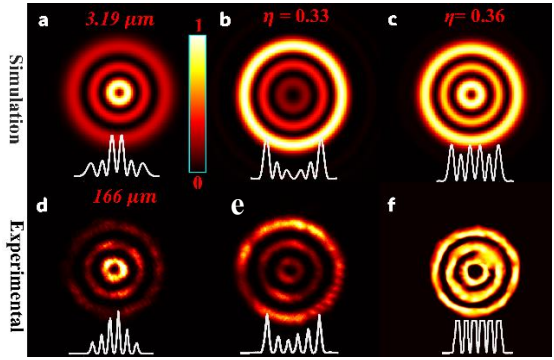

***Figure 5*** *The first row presents the simulated high-order R-TEM$_{21}$ mode profiles, while the second row shows the corresponding experimental results. The first column displays the unmodulated beam profile, whereas the second and third columns show the modulated profiles obtained at η = 0.45 and η = 0.55, respectively, demonstrating opposite and equal intensity redistribution among the rings. The scale and color bar provided in Figure 5(a, d) apply to the corresponding panels.*

modes of a RPCV beam into equal and opposite intensity profiles can be achieved by modulating the phase of the beam cross-section using the binary mask function by variation of the optimal tuning parameter η across different beam modes. The optimal values of η for different modes were determined empirically through systematic simulations, as η ranges between 0 and 1 and is easily tunable. The binary phase mask, determines the radial position at which a π phase shift is introduced, effectively dividing the aperture into two concentric regions with opposite phase. The resulting phase difference between the inner and outer zones, together with the chosen value of η, plays a crucial role in shaping the constructive and destructive interference that governs the energy redistribution in high-order R-TEM laser modes. Due to the presence of the grating function, the resulting energy modulation of the R-TEM modes appears in the first diffraction order of the computed DOE pattern. Furthermore, for a fixed mask diameter, the parameter η is a key factor in controlling how the phase mask shapes the optical energy distribution within specific diffraction orders. To analyze the controllable modulation mechanism, it is necessary to separately examine the diffraction contributions arising from the annular phase mask together with the beam function. By tuning different values of η, distinct diffraction patterns emerge.

These patterns directly influence and accurately regulate the modes. Consequently, each numerical value of η corresponds uniquely to a specific energy modulation associated with different high-order R-TEM laser modes. For other η values, either complete or partial destructive or constructive interference occurs at different spatial locations in the focal region, resulting in variations in the profile shape. Any desired intensity ratio among the rings can be achieved by precisely selecting the value of η to the required decimal accuracy.

**Conclusion**

In conclusion, we experimentally established a method that modulates the optical energy among the various concentric dark-core intensity rings and lobes of high-order standard R-TEM laser modes in a controllable manner. For a certain numerical value of η, all the low-intensity dark-core rings are swapped to highly intense rings and, and simultaneously, all the highly intense rings are switched to low-intensity rings, with a switchable intensity ratio among the rings of high-order R-TEM laser modes. This enables the creation of opposite intensity profile having more intense outer ring than the inner ring in one case and equal intensity profile having equal peak intensities in both rings in the second case, compared to the profile of standard high-order R-TEM laser modes.

Therefore, we anticipate that in many applications where highly intense lobes have a dominant effect, such as particle manipulation of micro- and nanoparticles, optical lithography, quantum information processing including quantum communication, and quantum cryptography, experimentally generated modulated high-order R-TEM laser modes may outperform standard high-order R-TEM laser modes. Also, it may find application in a near-field optical storage system for improving the recording ability [44].

Other high-order R-TEM laser modes can be used with this controllable technique. The creation of THz radiation and electron acceleration from the laser–plasma interaction mechanism is significantly impacted by a laser's energy redistribution [45,46]. Finally, this technique of modifying optical energy between rings can be applied in several fields, including plasmonic, nonlinear frequency conversion, and time-domain modulation of light pulses.